\documentclass[conference]{IEEEtran}
\usepackage{bm}
\usepackage{amsmath, amssymb}
\usepackage[only, boxast]{stmaryrd} 
\usepackage{siunitx}
\usepackage{pgfplots}
\usepackage{cancel}
\usepackage[inline]{enumitem}
\usepackage{booktabs}
\usepackage{mathtools}
\usepackage{xcolor}

\pgfplotsset{compat = newest}
\usetikzlibrary{shapes, calc}

\ifCLASSOPTIONcompsoc
\usepackage[caption=false,font=normalsize,labelfont=sf,textfont=sf,labelformat=parens]{subfig}
\else
\usepackage[caption=false,font=footnotesize,labelformat=parens,subrefformat=parens]{subfig}
\fi

\allowdisplaybreaks

\newcommand{\e}{\mathop{}\!\mathrm{e}}

\newcommand{\imj}{\mathrm{j}}
\newcommand{\diff}{\mathop{}\!\mathrm{d}}

\title{On Code Design for Wireless Channels with Additive Radar Interference}
\author{Federico Brunero, Daniela Tuninetti, and Natasha Devroye\\
University of Illinois at Chicago, Chicago, IL 60607, USA\\
Email: \{fbrune3, danielat, devroye\}@uic.edu}

\begin{document}
\maketitle

\begin{abstract}
This paper considers the problem of code design for a channel where communications and radar systems coexist, modeled as having both Additive White Gaussian Noise (AWGN) and Additive Radar Interference (ARI).
The issue of how to adapt or re-design convolutional codes (decoded by the Viterbi algorithm) and LDPC codes (decoded by the sum-product algorithm and optimized by using the EXIT chart method) to effectively handle the overall non-Gaussian ARI noise is investigated. A decoding metric is derived from the non-Gaussian ARI channel transition probability as a function of the Signal-to-Noise Ratio (SNR) and Interference-to-Noise Ratio (INR).

Two design methodologies are benchmarked against a baseline ``unaltered legacy system'', where a code designed for AWGN-only noise, but used on the non-Gaussian ARI channel, is decoded by using the AWGN-only metric (i.e., as if INR is zero).
The methodologies are:
M1) codes designed for AWGN-only noise, but decoded with the new metric that accounts for both SNR and INR; and
M2) codes optimized for the overall non-Gaussian ARI channel.
Both methodologies give better average Bit Error Rate (BER) in the high INR regime compared to the baseline. In the low INR regime, both methodologies perform as the baseline since in this case the radar interference is weak.  
Interestingly, the performance improvement of M2 over M1 is minimal. In practice, this implies that specifications in terms of channel error correcting codes for commercially available wireless systems need not be changed, and that it suffices to use an appropriate INR-based decoding metric in order to effectively cope with the ARI.
\end{abstract}

\section{Introduction}
As the demand for wireless services increases, one of several solutions have been proposed to open up spectrum is to allow communications and radar systems to share frequency bands, as surveyed in~\cite{7852330,8332962} and references therein.

\paragraph*{Past Work}
From the perspective of a communication system only (where the communication system seeks to adapt to the unalterable and uncooperative radar system), the authors in~\cite{7852330} investigated the {\it Shannon capacity} of the AWGN channel suffering from an additive constant-modulo interference caused by a co-existing radar transmission; results about the capacity achieving channel input distribution were obtained when the radar interference is larger than the signal of interest. In this case the (complex-valued) channel ``loses'' one of the two real-valued dimensions. In~\cite{8332962}, the Symbol Error Rate (SER) for the channel model in~\cite{7852330} was investigated for commercially employed {\it uncoded} modulation systems; the authors derived decoding regions for optimal and suboptimal detection schemes, and designed optimal signal constellations such that either the transmission rate was maximized or the SER was minimized, including extensions to OFDM systems.

To the best of our knowledge powerful {\it error correcting codes} for the channel model in~\cite{7852330} have not been investigated.
LDPC codes are a widely studied class of powerful error correcting codes. The authors in~\cite{910578} examined the design of capacity-approaching irregular LDPC codes, focusing mainly on 
channels with binary inputs. In~\cite{910580} the so-called ``Gaussian approximation'' was proposed to simplify the density evolution analysis of LDPC codes and it has become the method of choice also for the so-called ``EXIT chart method''~\cite[Chapter 4, p. 238]{richardson_urbanke_2008}. 
However, in spite of the wide literature on the design of LDPC codes, we are not aware of any LDPC code designed for the communications and radar channel model of \cite{7852330, 8332962}.
This paper takes the first step towards understanding the performance of practical \emph{coded} systems for wireless channels that suffer from additive radar interference.

\paragraph*{Contributions}
In this paper we study the problem of code design for the channel model from~\cite{7852330} that is characterized by two parameters: Signal-to-Noise Ratio (SNR) for the signal of interest, and Interference-to-Noise Ratio (INR) for the nuisance radar signal. Based on the channel transition probability, we derive a decoding metric that is a function of SNR and INR, and that reduces to the AWGN-only (i.e., related to the Euclidean distance) one when INR is zero. 

Our baseline is an ``unaltered legacy system'', where a code designed for AWGN-only noise, but used on the radar interfered non-AWGN channel, is decoded using the AWGN-only metric (i.e., as if INR is zero); we shall refer to this baseline as \textbf{M0}.
We explore the following two design methodologies.
\textbf{M1}: codes designed for AWGN-only noise, but decoded with the new metric that accounts for both SNR and INR.
\textbf{M2}: codes optimized for the overall non-Gaussian channel.
Intuitively, M2 outperforms M1, which outperforms M0, for ``reasonably'' optimized codes. 
Moreover, it is expected that M0, M1 and M2 give essentially the same performance when INR is negligible with respect to the SNR.

For M1, we first analyze convolutional codes decoded by the Viterbi algorithm~\cite{lin2004error}, and then LDPC codes decoded by the sum-product algorithm~\cite{richardson_urbanke_2008}.
For M2, we design LDPC codes by the EXIT chart method with Gaussian approximations~\cite[Chapter 4, p. 238]{richardson_urbanke_2008} and some optimization tricks~\cite{Amraoui:85786}.

Both M1 and M2 give better average Bit Error Rate (BER) compared to M0 when $\text{INR} \gg \text{SNR}$.
When $\text{INR} \ll \text{SNR}$, M1 and M2 perform as M0 since in this case the radar interference is weak.
Interestingly, we observe that M2 only offers a small improvement over M1 in general.
This may be intuitively understood as follows: AWGN is known to be the ``worst noise'' in terms of capacity among all noises with the same second-moment; thus, a code designed for AWGN-only noise is intrinsically robust and performs quite well in non-AWGN scenarios. In practice, this may imply that current wireless system specifications in terms of forward error correcting channel codes need not be changed, and that in order to effectively cope with the additional radar interference it suffices to use the appropriate INR-based decoding metric with the existing codes.

\paragraph*{Paper Organization}
The paper is organized as follows. The channel model, the modified decoding algorithms and the LDPC code design are introduced in Section~\ref{sec: Channel Model and Codes Used}. Section~\ref{sec: Numerical Results} presents simulation results. Section~\ref{sec: Conclusion} concludes the paper.

\section{Channel Model and Codes Used}
\label{sec: Channel Model and Codes Used}
In the following capital letters represent random variables and lower case letters represent their realizations. The codeword $\bm{x} = (x_{0}, x_{1}, \cdots, x_{N - 1})$ and the complex-valued received sequence $\bm{y} = (y_{0}, y_{1}, \cdots, y_{N - 1})$ are sequences of length $N$.
In this paper, for sake of space, we only report results for the case where the coded bits are mapped to the BPSK constellation, thus the coded sequence/codeword and the channel output sequence have the same length.

\paragraph{Channel Model}
\label{par:ChModel}
Consider the following AWGN channel model with Additive Radar Interference (ARI) from~\cite{7852330}
\begin{align}
\label{eqn: AWGN+ARI channel model}
  Y = \sqrt{S}X + \sqrt{I}\e^{\imj\Theta} + Z,
\end{align}
where 
$X \in \{-1, +1\}$ is the transmitted symbol from the BPSK constellation, 
$\Theta$ is the radar phase uniformly distributed in $[0, 2\pi)$, 
$Z \sim \mathcal{N}(0, 1)$ is the complex-valued Gaussian noise, and 
$Y$ is the channel output.
The random variables $(X, \Theta, Z)$ are mutually independent. 
The pair $(S, I)\in\mathbb{R}^2_+$ is fixed and known at the communications receiver, where $S$ is the SNR and $I$ the INR.
Let $W \coloneqq \sqrt{I}\e^{\imj\Theta} + Z$, its density is given by 
\begin{align}
\label{eqn: AWGN+ARI noise transition probability}
  f_{W}(w) = \frac{\e^{-|w|^{2} - I}}{\pi}I_{0}\Big(2\sqrt{I|w|^{2}}\Big),
\end{align}
where $I_{0}$ is the modified Bessel function of the first kind of order zero.
The channel transition probability for~\eqref{eqn: AWGN+ARI channel model} is simply
\begin{align}
\label{eqn: AWGN+ARI channel transition probability}
   f_{Y \mid X}(y \mid x) = f_{W}\big(y - \sqrt{S}x\big). 
\end{align}
The channel is memoryless.
When designing coded schemes for the memoryless channel in~\eqref{eqn: AWGN+ARI channel model} with equally likely codewords, one needs the following log-likelihood function of the received sequence $\bm{y}$ given the transmitted codeword $\bm{x}$
\begin{subequations}
  \begin{align}
    &M(\bm{y} \mid \bm{x}) 
      = \ln{f_{Y \mid X}(\bm{y}\mid\bm{x})} 
      = \ln{\prod_{\ell = 0}^{N - 1}}{f_{Y \mid X}(y_\ell \mid x_\ell)}
      \notag
    \\&= \sum_{\ell = 0}^{N - 1}\ln\Big(I_{0}\Big(2\sqrt{I|y_\ell - \sqrt{S}x_\ell|^{2}}\Big)\Big) 
    \label{eqn: codewordLLR: INR term}
    \\&- \sum_{\ell = 0}^{N - 1}|y_\ell - \sqrt{S}x_\ell|^{2} - \alpha, \quad \alpha \coloneqq N(\ln(\pi) + I) > 0.
    \label{eqn: codewordLLR: AWGN-only term}
  \end{align}
  \label{eqn: codewordLLR}%
\end{subequations}
Note that the ``metric'' in~\eqref{eqn: codewordLLR} reduces to the standard Euclidean distance (i.e., the term in~\eqref{eqn: codewordLLR: AWGN-only term}) when INR is zero, because in this case the term in~\eqref{eqn: codewordLLR: INR term} is zero;
in this case, one can further manipulate~\eqref{eqn: codewordLLR} to reduce it to ``correlation metric'' where the received signal is projected over the possible inputs.

\paragraph{Baseline M0}
\label{par:M0}
Our baseline schemes are codes designed for AWGN-only noise, used on the AWGN+ARI channel, but decoded using the AWGN-only metric. We look at two cases:
\begin{enumerate*}[label=\alph*)]
\item\label{M0-a} an unaltered legacy system, where the decoding metric is as if the channel was $Y = \sqrt{S}X + Z$ (i.e., as if INR is zero), and
\item\label{M0-b} the ARI is treated as a Gaussian noise, where the decoding metric is as if the channel was $Y = \sqrt{S/(1+I)}X + Z$.
\end{enumerate*}

\paragraph{Methodology M1}
\label{par:M1}
Here, codes designed for AWGN-only noise are decoded with the new metric in~\eqref{eqn: codewordLLR}.
We consider convolutional codes and LDPC codes. 
The Viterbi algorithm, the optimal maximum a posteriori sequence detector, is used for decoding convolutional codes.
The Sum-Product Algorithm (SPA), an approximation of the optimal maximum a posteriori symbol-by-symbol detector, is used for decoding LDPC codes.
In both cases (since we use BPSK modulation), the decoding metric used is a function of the log-likelihood in~\eqref{eqn: codewordLLR}.

\paragraph{Methodology M2}
\label{par:M2}
Here, we optimize codes for the overall AWGN+ARI channel. We only consider LDPC codes with the SPA decoder.
The SPA is one of the most efficient iterative algorithms for decoding LDPC codes; it is essentially based on the computation of marginal \emph{a posteriori bit probabilities}.
During the initialization step, we assign to each variable node the a posteriori Log-Likelihood Ratio (LLR) defined as
\begin{subequations}
\begin{align}
  \mathrm{LLR}(y_{\ell})
  &\coloneqq \ln{\frac{f_{X \mid Y}(x_{\ell} = -1 \mid y_{\ell})}{f_{X \mid Y}(x_{\ell} = +1 \mid y_{\ell})}}
  \\&= -4\Re(y_{\ell})\sqrt{S}
     -\ln\frac{I_{0}\Big(2\sqrt{I|y_{\ell} - \sqrt{S}|^{2}}\Big)}{I_{0}\Big(2\sqrt{I|y_{\ell} + \sqrt{S}|^{2}}\Big)},
\end{align}
\label{eqn: LLR for SPA in the log-domain}%
\end{subequations}
derived from (recall that we assume equally likely symbols)
\begin{align}
  f_{X \mid Y}(x_{\ell} \mid y_{\ell}) = \frac{f_{W}\big(y_{\ell} - \sqrt{S}x_{\ell}\big)}{f_{W}\big(y_{\ell} - \sqrt{S}\big) + f_{W}\big(y_{\ell} + \sqrt{S}\big)}.
\end{align}

We optimize our LDPC codes based on the EXIT method with customary Gaussian approximations, which is used as a proxy for exact density evolution in the limit for infinite iterations and infinite code-length.
EXIT charts help to visualize the asymptotic performance under iterative Belief Propagation (BP) decoding. Originally introduced in the context of turbo codes and based on the extrinsic information exchanged between variable and check nodes, this technique has been successfully used for the design of LDPC codes as well~\cite{richardson_urbanke_2008}.

\smallskip
In order to use the EXIT chart method, it is convenient to introduce the \emph{degree distributions from an edge perspective}, i.e., the polynomials $\lambda(x) = \sum_{i = 2}^{d_{\text{v}}}{\lambda_{i}x^{i-1}}$ and $\rho(x) = \sum_{i = 2}^{d_{\text{c}}}{\rho_{i}x^{i-1}}$, where the coefficients $\lambda_{i}$ ($\rho_{i}$) represent the fraction of edges that are connected to variable (check) nodes of degree $i$, and  $d_{\text{v}}$ ($d_{\text{c}}$) is the maximum degree of the variable (check) degree distribution. The pair $(\lambda, \rho)$ represents an LDPC \emph{ensemble} with \emph{design rate} given by
\begin{align}
\label{eqn: Design rate of the LDPC code}
    r(\lambda, \rho) = 1 - \frac{\int_{0}^{1}{\rho(x)\diff{x}}}{\int_{0}^{1}{\lambda(x)\diff{x}}} = 1 - \frac{\sum_{i}{\rho_{i}/i}}{\sum_{i}{\lambda_{i}/i}}.
\end{align}

Consider our ARI channel with BPSK modulation\footnote{We consider this mapping to be consistent with all the derivations in~\cite{richardson_urbanke_2008} concerning entropy and capacity associated with a symmetric $L$-density; the standard BPSK mapping is used in computer simulations, but the different convention does not affect the results.} $0 \to +1$ and $1 \to -1$. With the definition of LLR in~\eqref{eqn: LLR for SPA in the log-domain}, we denote the conditional density of the random variable $\mathrm{LLR}(Y)$ on $X = +1$ as $\mathsf{a}_{\text{AWGN+ARI}}$, and refer to it as \emph{$L$-density}.
The \emph{entropy} $\mathsf{H}$ and the \emph{capacity} $\mathsf{C}$ of the channel 
are given by
\begin{align}
  \mathsf{H}
  &= \int_{-\infty}^{+\infty}{\mathsf{a}_{\text{AWGN+ARI}}(u)\log_{2}\left(1 + \e^{-u}\right)\diff{u}}
\label{eqn: Hdef}
=1-\mathsf{C}.
\end{align}

Referring to the EXIT chart method in~\cite[Definition 4.136, p. 237]{richardson_urbanke_2008} and using the Gaussian approximation for the ``intermediate'' densities, the functions
\begin{align}
    v_{\hat{\mathsf{h}}}(\mathsf{h}) &\coloneqq \sum_{i}\lambda_{i}\mathsf{H}(\mathsf{a}_{\text{AWGN+ARI}(\hat{\mathsf{h}})} \circledast \mathcal{N}(m_{i}, 2m_{i})),
\label{eqn: vdef}
    \\
  c(\mathsf{h}) &\coloneqq 1 - \sum_{i}\rho_{i}\psi\left((i - 1)\psi^{-1}(1 - \mathsf{h})\right),
\label{eqn: cdef}
\end{align}
are the entropies at the output of the variable and check nodes, respectively, as a function of the input entropy $\mathsf{h}$, when the transmission takes place over the AWGN+ARI channel; the parameter $m_{i} \coloneqq (i - 1)\psi^{-1}(\mathsf{h})$ is the mean of the Gaussian density, the function $\psi(m)$ is the entropy of a Gaussian density with mean $m$ and variance $2m$; the term $\mathsf{a}_{\text{AWGN+ARI}(\hat{\mathsf{h}})}$ is the $L$-density for the AWGN+ARI channel with parameters $(S, I)$ such that entropy in~\eqref{eqn: Hdef} is equal to the \emph{channel parameter} $\hat{\mathsf{h}}$.

For the iterative algorithm to converge, the condition $v_{\hat{\mathsf{h}}}(c(\mathsf{h})) \leq \mathsf{h}$ must be satisfied among the quantities defined in~\eqref{eqn: vdef}-\eqref{eqn: cdef}. Therefore, if we fix the channel parameter $\hat{\mathsf{h}}$ and the $\rho$ distribution, we can find the coefficients $\lambda_{i}$ by solving the following linear program
\begin{align}
    &\max_{\lambda_{i} \geq 0}
    \Big\{\sum_{i \geq 2}\frac{\lambda_{i}}{i} :
    \sum_{i \geq 2}{\lambda_{i}} = 1,
    \sum_{i \geq 2}\lambda_{i} \xi_{i, c} \leq \mathsf{h}, 
    \mathsf{h} \in [0, 1]
    \Big\}, 
\label{eqn: optimizationL}
\\
    &\xi_{i, c} \coloneqq \mathsf{H}(\mathsf{a}_{\text{AWGN+ARI}(\hat{\mathsf{h}})} \circledast \mathcal{N}(m_{i, c}, 2m_{i, c})), 
\label{eqn: xidef}
\\ 
    &m_{i, c} \coloneqq (i - 1)\psi^{-1}(c(\mathsf{h})),
\end{align}
where the objective function in~\eqref{eqn: optimizationL} is equivalent to maximization of the code rate in~\eqref{eqn: Design rate of the LDPC code}. Similarly, keeping fixed the $\lambda$ distribution, we can solve the linear program
\begin{align}
    &\min_{\rho_{i} \geq 0}
    \Big\{\sum_{i \geq 2}\frac{\rho_{i}}{i} : 
    \sum_{i \geq 2}{\rho_{i}} = 1,
    \sum_{i \geq 2} \rho_{i} \nu_{i,v} \geq 1 - \mathsf{h},
    \mathsf{h} \in [0, 1]
    \Big\},
\label{eqn: optimizationR}
\\
    &\nu_{i,v}\coloneqq \psi\left((i - 1)\psi^{-1}(1 - v_{\hat{\mathsf{h}}}(\mathsf{h}))\right),
\label{eqn: nudef}
\end{align}
to find the coefficients $\rho_{i}$, which maximize the code rate too.

We aim to design LDPC codes for the AWGN+ARI channel that outperform LDPC codes optimized for the AWGN-only channel. Our optimization algorithm has these steps:
\begin{enumerate*}
\item initialization: we choose $\rho(x) = (1 - \rho)x^{d_{\text{c}} - 1} + \rho x^{d_{\text{c}}}$ with the coefficient $\rho$ uniformly distributed on $[1/2,1]$,
equal for both the AWGN-only channel and the AWGN+ARI channel optimization\footnote{We use a $\rho(x)$ polynomial with coefficients $\rho_{i}$ concentrated at two consecutive degrees since this choice makes easier the optimization~\cite{910580}.};
\item we 
optimize $\lambda(x)$ given the initial $\rho(x)$ by solving~\eqref{eqn: optimizationL}, and then we optimize $\rho(x)$ given the found $\lambda(x)$ by solving~\eqref{eqn: optimizationR};
\item we choose the pair $(S, I)$ for AWGN+ARI channel, and $S$ for the AWGN-only channel, such that the design rate, after the optimization of the degree distributions, is approximately the same. Here, we report results for rate around one-half.
\end{enumerate*}

\section{Simulation Results}
\label{sec: Numerical Results}

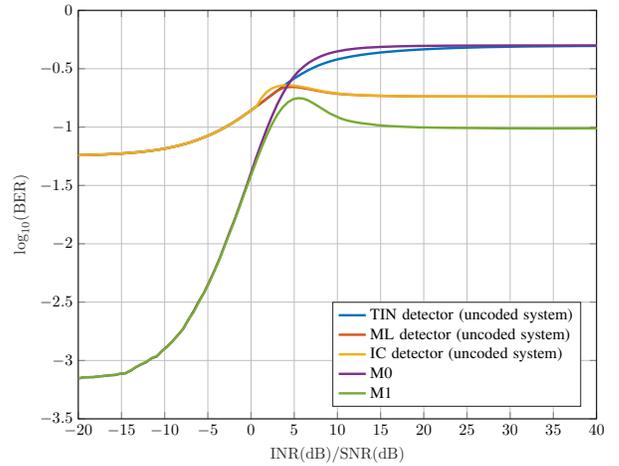
\begin{figure}
  \centering
%
%
\definecolor{mycolor1}{rgb}{0.00000,0.44700,0.74100}%
\definecolor{mycolor2}{rgb}{0.85000,0.32500,0.09800}%
\definecolor{mycolor3}{rgb}{0.92900,0.69400,0.12500}%
\definecolor{mycolor4}{rgb}{0.49400,0.18400,0.55600}%
\definecolor{mycolor5}{rgb}{0.46600,0.67400,0.18800}%
\begin{tikzpicture}[scale = 0.6]

\begin{axis}[%
width=4.521in,
height=3.566in,
at={(0.758in,0.481in)},
scale only axis,
xmin=-20,
xmax=40,
xlabel style={font=\color{white!15!black}},
xlabel={$\mathrm{INR}(\si{\decibel})/\mathrm{SNR}(\si{\decibel})$},
ymin=-3.5,
ymax=0,
ylabel style={font=\color{white!15!black}},
ylabel={$\log_{10}(\mathrm{BER})$},
axis background/.style={fill=white},
xmajorgrids,
ymajorgrids,
legend style={at={(0.97,0.03)}, anchor=south east, legend cell align=left, align=left, draw=white!15!black}
]
\addplot [color=mycolor1, line width=1.5pt]
  table[row sep=crcr]{%
-20	-1.24146657776271\\
-16.969696969697	-1.23454007588517\\
-15.1515151515152	-1.22766362141404\\
-12.7272727272727	-1.21328406087925\\
-10.9090909090909	-1.19639268267739\\
-9.6969696969697	-1.18055477896745\\
-8.48484848484848	-1.16089921729284\\
-7.27272727272727	-1.13658196247791\\
-6.06060606060606	-1.10603860820657\\
-4.84848484848485	-1.07006023334952\\
-4.24242424242424	-1.04951831080761\\
-3.63636363636364	-1.02694145424127\\
-3.03030303030303	-1.00276656312387\\
-2.42424242424243	-0.977052270868732\\
-1.81818181818182	-0.949334159965403\\
-1.21212121212121	-0.920135278239329\\
0	-0.857022694413203\\
1.21212121212121	-0.790729009543718\\
4.24242424242424	-0.621800211589445\\
4.84848484848485	-0.59069596065757\\
5.45454545454545	-0.562235172266675\\
6.06060606060606	-0.535634171085455\\
6.66666666666667	-0.511194139950874\\
7.27272727272727	-0.489215121750036\\
7.87878787878788	-0.469845142775078\\
8.48484848484848	-0.45307379688483\\
9.09090909090909	-0.438359662154888\\
9.6969696969697	-0.425678141739184\\
10.3030303030303	-0.414561831829822\\
11.5151515151515	-0.396225761689401\\
12.7272727272727	-0.381525144001209\\
13.9393939393939	-0.36965655107754\\
15.1515151515151	-0.359797913149293\\
16.969696969697	-0.347956828652976\\
18.7878787878788	-0.33854416913811\\
21.2121212121212	-0.329027219580176\\
23.6363636363636	-0.322080212484202\\
27.2727272727273	-0.314830832842496\\
31.5151515151515	-0.309605430716701\\
37.5757575757576	-0.305567675899233\\
40	-0.304566483834272\\
};
\addlegendentry{TIN detector (uncoded system)}

\addplot [color=mycolor2, line width=1.5pt]
  table[row sep=crcr]{%
-20	-1.24146657776271\\
-16.969696969697	-1.23454007588517\\
-15.1515151515152	-1.22766362141404\\
-12.7272727272727	-1.21328406087925\\
-10.9090909090909	-1.19639268267739\\
-9.6969696969697	-1.18055477896745\\
-8.48484848484848	-1.16089921729284\\
-7.27272727272727	-1.13658196247791\\
-6.06060606060606	-1.10603860820657\\
-4.84848484848485	-1.07006023334952\\
-4.24242424242424	-1.04951831080761\\
-3.63636363636364	-1.02694145424127\\
-3.03030303030303	-1.00276656312387\\
-2.42424242424243	-0.977052270868732\\
-1.81818181818182	-0.949334159965403\\
-1.21212121212121	-0.920135278239329\\
0	-0.857022694413203\\
1.21212121212121	-0.790729009543718\\
3.03030303030303	-0.688744274287366\\
3.63636363636364	-0.667319210784754\\
4.24242424242424	-0.658130409650681\\
4.84848484848485	-0.656069778369144\\
5.45454545454545	-0.659421056662907\\
6.66666666666667	-0.67434479372303\\
8.48484848484848	-0.699671956903686\\
9.09090909090909	-0.706694912991487\\
10.3030303030303	-0.716734964011323\\
11.5151515151515	-0.722890683829974\\
13.9393939393939	-0.730643523374916\\
18.7878787878788	-0.737157509702264\\
31.5151515151515	-0.738582478520328\\
40	-0.738161626309385\\
};
\addlegendentry{ML detector (uncoded system)}

\addplot [color=mycolor3, line width=1.5pt]
  table[row sep=crcr]{%
-20	-1.24146657776271\\
-16.969696969697	-1.23454007588517\\
-15.1515151515152	-1.22766362141404\\
-12.7272727272727	-1.21328406087925\\
-10.9090909090909	-1.19639268267739\\
-9.6969696969697	-1.18055477896745\\
-8.48484848484848	-1.16089921729284\\
-7.27272727272727	-1.13658196247791\\
-6.06060606060606	-1.10603860820657\\
-4.84848484848485	-1.07006023334952\\
-4.24242424242424	-1.04951831080761\\
-3.63636363636364	-1.02694145424127\\
-3.03030303030303	-1.00276656312387\\
-2.42424242424243	-0.977052270868732\\
-1.81818181818182	-0.949334159965403\\
-1.21212121212121	-0.920135278239329\\
0	-0.857022694413203\\
0.606060606060609	-0.824531676245407\\
1.21212121212121	-0.751070866281232\\
1.81818181818182	-0.698368922631644\\
2.42424242424243	-0.668872459972746\\
3.03030303030303	-0.651804211145524\\
3.63636363636364	-0.642976603329259\\
4.24242424242424	-0.640587639298488\\
4.84848484848485	-0.643631816471732\\
5.45454545454545	-0.649608533074968\\
6.06060606060606	-0.657646416978295\\
8.48484848484848	-0.695556868059676\\
9.6969696969697	-0.708988768177441\\
10.9090909090909	-0.717769993891416\\
12.1212121212121	-0.723733557384698\\
13.9393939393939	-0.729353766273341\\
18.7878787878788	-0.736669346614917\\
26.0606060606061	-0.738297110472672\\
32.7272727272727	-0.738392212322211\\
36.969696969697	-0.738228174651823\\
40	-0.738278092601561\\
};
\addlegendentry{IC detector (uncoded system)}

\addplot [color=mycolor4, line width=1.5pt]
  table[row sep=crcr]{%
-20	-3.15428198203334\\
-19.3939393939394	-3.1444808443322\\
-18.1818181818182	-3.14206473528057\\
-17.5757575757576	-3.13608262304214\\
-16.969696969697	-3.13253251214095\\
-16.3636363636364	-3.12609840213554\\
-15.7575757575758	-3.12436006299583\\
-15.1515151515152	-3.11350927482752\\
-14.5454545454545	-3.10957898119909\\
-13.9393939393939	-3.08724669632868\\
-13.3333333333333	-3.05749589383192\\
-12.7272727272727	-3.03385826726097\\
-12.1212121212121	-3.01412464269161\\
-11.5151515151515	-2.98088370955293\\
-10.9090909090909	-2.96577273922945\\
-10.3030303030303	-2.91793306571488\\
-9.6969696969697	-2.87909718238547\\
-9.09090909090909	-2.83505262737816\\
-7.87878787878788	-2.73165608604894\\
-7.27272727272727	-2.65364702554936\\
-6.66666666666666	-2.57954914089393\\
-6.06060606060606	-2.49457867241672\\
-5.45454545454545	-2.420445039599\\
-4.84848484848485	-2.32753268693192\\
-4.24242424242424	-2.2298847052129\\
-3.63636363636364	-2.1253442674019\\
-2.42424242424243	-1.88917468994504\\
-1.81818181818182	-1.77640023535031\\
-0.606060606060609	-1.52219888301185\\
0.606060606060609	-1.25937681195837\\
1.21212121212121	-1.13544145765727\\
1.81818181818182	-1.01609870169747\\
2.42424242424243	-0.906693908590057\\
3.03030303030303	-0.805441259673259\\
3.63636363636364	-0.71891433865953\\
4.24242424242424	-0.641893754793585\\
4.84848484848485	-0.577569397821399\\
5.45454545454545	-0.524865111446452\\
6.06060606060606	-0.483955385025695\\
6.66666666666667	-0.448822131282867\\
7.27272727272727	-0.422743794129325\\
7.87878787878788	-0.400051461330918\\
8.48484848484848	-0.381837850557964\\
9.09090909090909	-0.367333203675841\\
9.6969696969697	-0.35595452149758\\
10.9090909090909	-0.338679604550308\\
12.1212121212121	-0.327464372169842\\
13.3333333333333	-0.319581081040965\\
15.1515151515151	-0.312779702637336\\
16.969696969697	-0.308180569200367\\
19.3939393939394	-0.304522701500439\\
23.6363636363636	-0.302052659409981\\
30.9090909090909	-0.300628892843683\\
40	-0.300666209293354\\
};
\addlegendentry{M0}

\addplot [color=mycolor5, line width=1.5pt]
  table[row sep=crcr]{%
-20	-3.15428198203334\\
-19.3939393939394	-3.1444808443322\\
-18.1818181818182	-3.14206473528057\\
-17.5757575757576	-3.13608262304214\\
-15.7575757575758	-3.12436006299583\\
-15.1515151515152	-3.11407366019857\\
-14.5454545454545	-3.10957898119909\\
-13.9393939393939	-3.08618614761628\\
-13.3333333333333	-3.05700040663396\\
-12.1212121212121	-3.01010543628123\\
-11.5151515151515	-2.98213228103649\\
-10.9090909090909	-2.96297212024422\\
-10.3030303030303	-2.91865269219587\\
-9.6969696969697	-2.88239730830992\\
-8.48484848484848	-2.78462684721658\\
-7.87878787878788	-2.73095429034237\\
-7.27272727272727	-2.64975198166584\\
-6.66666666666666	-2.58787559382668\\
-6.06060606060606	-2.50182733936346\\
-5.45454545454545	-2.42101715729721\\
-4.84848484848485	-2.33292079453578\\
-4.24242424242424	-2.23232477597204\\
-3.63636363636364	-2.13018179202067\\
-3.03030303030303	-2.01291497037588\\
-2.42424242424243	-1.89928491342692\\
-1.81818181818182	-1.7813857302549\\
-1.21212121212121	-1.65925852501002\\
-0.606060606060609	-1.5341266134284\\
0	-1.41473107923661\\
0.606060606060609	-1.29319690296266\\
1.21212121212121	-1.17753769424133\\
1.81818181818182	-1.07120486236776\\
2.42424242424243	-0.975937339106345\\
3.03030303030303	-0.895349034891318\\
3.63636363636364	-0.832777215838796\\
4.24242424242424	-0.788759283170855\\
4.84848484848485	-0.761630179466657\\
5.45454545454545	-0.751913880809575\\
6.06060606060606	-0.755908514799145\\
6.66666666666667	-0.773260936662354\\
7.27272727272727	-0.79874220002884\\
9.09090909090909	-0.882168792397302\\
10.3030303030303	-0.923643596969903\\
10.9090909090909	-0.938966183473539\\
11.5151515151515	-0.950057465320192\\
12.1212121212121	-0.958962792132972\\
12.7272727272727	-0.965877110887632\\
13.9393939393939	-0.976850465946079\\
14.5454545454545	-0.983004245503828\\
15.7575757575758	-0.990264483882832\\
17.5757575757576	-0.996854100654993\\
19.3939393939394	-1.00225549219751\\
22.4242424242424	-1.00606582891165\\
27.8787878787879	-1.01003876363848\\
29.6969696969697	-1.01115131287357\\
31.5151515151515	-1.01138753991019\\
33.9393939393939	-1.01193626054727\\
40	-1.01092412131047\\
};
\addlegendentry{M1}

\end{axis}
\end{tikzpicture}%
  \caption{\small BER vs INR for BPSK modulation in AWGN+ARI channel with $S = \SI{1}{\decibel}$.
  }
  \label{fig: Bit Error Rate (BER) vs I for BPSK modulation in AWGN+ARI channel with S = 1dB (convolutional code)}
\end{figure}

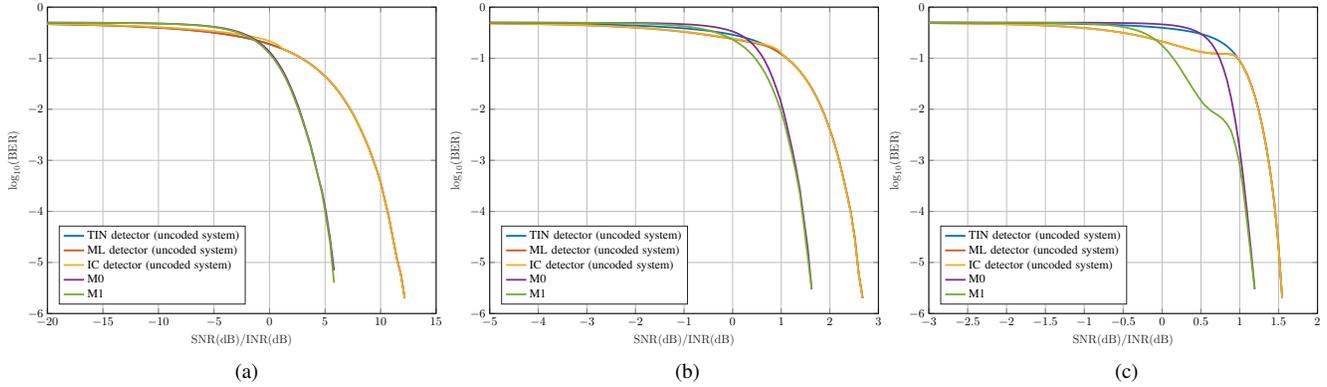
\begin{figure*}
  \centering
  \subfloat[\label{fig: I = 1dB}]{
%
%
\definecolor{mycolor1}{rgb}{0.00000,0.44700,0.74100}%
\definecolor{mycolor2}{rgb}{0.85000,0.32500,0.09800}%
\definecolor{mycolor3}{rgb}{0.92900,0.69400,0.12500}%
\definecolor{mycolor4}{rgb}{0.49400,0.18400,0.55600}%
\definecolor{mycolor5}{rgb}{0.46600,0.67400,0.18800}%
\begin{tikzpicture}[trim axis left, scale = 0.45]

\begin{axis}[%
width=4.521in,
height=3.566in,
at={(0.758in,0.481in)},
scale only axis,
xmin=-20,
xmax=15,
xlabel style={font=\color{white!15!black}},
xlabel={$\mathrm{SNR}(\si{\decibel})/\mathrm{INR}(\si{\decibel})$},
ymin=-6,
ymax=0,
ylabel style={font=\color{white!15!black}},
ylabel={$\log_{10}(\mathrm{BER})$},
axis background/.style={fill=white},
xmajorgrids,
ymajorgrids,
legend style={at={(0.03,0.03)}, anchor=south west, legend cell align=left, align=left, draw=white!15!black}
]
\addplot [color=mycolor1, line width=1.5pt]
  table[row sep=crcr]{%
-20	-0.330791939178219\\
-17.1717171717172	-0.342618070426582\\
-14.6969696969697	-0.357089542924641\\
-12.5757575757576	-0.374153756750683\\
-10.8080808080808	-0.392367006134862\\
-9.39393939393939	-0.410297201642958\\
-7.97979797979798	-0.432625396349387\\
-6.91919191919192	-0.452930826985789\\
-5.85858585858586	-0.476540990817394\\
-4.7979797979798	-0.504602764728126\\
-3.73737373737374	-0.537943581711577\\
-3.03030303030303	-0.563966228683636\\
-2.32323232323232	-0.593294648764207\\
-1.61616161616162	-0.627029113634521\\
-0.90909090909091	-0.665608582622845\\
-0.202020202020201	-0.709934209640064\\
0.151515151515152	-0.734422825885726\\
0.858585858585858	-0.789483510565994\\
1.56565656565656	-0.853013530624708\\
1.91919191919192	-0.888371246857336\\
2.27272727272727	-0.92691614621663\\
2.62626262626263	-0.968046484847754\\
2.97979797979798	-1.01321931130067\\
3.33333333333333	-1.0617255102895\\
3.68686868686869	-1.11512537180182\\
4.04040404040404	-1.17204301767367\\
4.39393939393939	-1.23418248469008\\
4.74747474747475	-1.30248296704874\\
5.1010101010101	-1.37557350720414\\
5.45454545454545	-1.4551261567652\\
5.80808080808081	-1.54228458950412\\
6.16161616161616	-1.63764949240515\\
6.51515151515152	-1.74381161741023\\
6.86868686868687	-1.86015238535369\\
7.22222222222222	-1.98275791545235\\
7.57575757575757	-2.11815903167507\\
7.92929292929293	-2.26640153903866\\
8.28282828282828	-2.42805836492554\\
8.63636363636364	-2.60380065290427\\
8.98989898989899	-2.78489141894691\\
9.34343434343434	-3.00086945871263\\
9.6969696969697	-3.21395878975745\\
10.0505050505051	-3.47108329972234\\
10.7575757575758	-4.1249387366083\\
11.4646464646465	-4.92081875395237\\
11.8181818181818	-5.22184874961636\\
12.1717171717172	-5.69897000433602\\
};
\addlegendentry{TIN detector (uncoded system)}

\addplot [color=mycolor2, line width=1.5pt]
  table[row sep=crcr]{%
-20	-0.330965921389051\\
-16.8181818181818	-0.34478662703135\\
-14.3434343434343	-0.360316330024983\\
-12.2222222222222	-0.378169652328896\\
-10.4545454545455	-0.397214257891111\\
-9.04040404040404	-0.416298053532056\\
-7.62626262626263	-0.43975731562734\\
-6.21212121212121	-0.468453389388017\\
-5.15151515151515	-0.494769956015819\\
-4.09090909090909	-0.52591637630394\\
-3.03030303030303	-0.563966228683636\\
-2.32323232323232	-0.593294648764207\\
-1.61616161616162	-0.627029113634521\\
-0.90909090909091	-0.665608582622845\\
-0.202020202020201	-0.709934209640064\\
0.151515151515152	-0.734422825885726\\
0.858585858585858	-0.789483510565994\\
1.56565656565656	-0.853013530624708\\
1.91919191919192	-0.888371246857336\\
2.27272727272727	-0.92691614621663\\
2.62626262626263	-0.968046484847754\\
2.97979797979798	-1.01321931130067\\
3.33333333333333	-1.0617255102895\\
3.68686868686869	-1.11512537180182\\
4.04040404040404	-1.17204301767367\\
4.39393939393939	-1.23418248469008\\
4.74747474747475	-1.30248296704874\\
5.1010101010101	-1.37557350720414\\
5.45454545454545	-1.4551261567652\\
5.80808080808081	-1.54228458950412\\
6.16161616161616	-1.63764949240515\\
6.51515151515152	-1.74381161741023\\
6.86868686868687	-1.86015238535369\\
7.22222222222222	-1.98275791545235\\
7.57575757575757	-2.11815903167507\\
7.92929292929293	-2.26640153903866\\
8.28282828282828	-2.42805836492554\\
8.63636363636364	-2.60380065290427\\
8.98989898989899	-2.78489141894691\\
9.34343434343434	-3.00086945871263\\
9.6969696969697	-3.21395878975745\\
10.0505050505051	-3.47108329972234\\
10.7575757575758	-4.1249387366083\\
11.4646464646465	-4.92081875395237\\
11.8181818181818	-5.22184874961636\\
12.1717171717172	-5.69897000433602\\
};
\addlegendentry{ML detector (uncoded system)}

\addplot [color=mycolor3, line width=1.5pt]
  table[row sep=crcr]{%
-20	-0.327573311615083\\
-16.8181818181818	-0.339980227820583\\
-13.989898989899	-0.355831328152536\\
-11.5151515151515	-0.375251880878363\\
-9.39393939393939	-0.397829277696751\\
-7.97979797979798	-0.41676380801437\\
-6.56565656565657	-0.439539793095722\\
-5.15151515151515	-0.467738780564485\\
-3.73737373737374	-0.502059098469381\\
-3.03030303030303	-0.522778869034827\\
-2.32323232323232	-0.546677072359426\\
-1.61616161616162	-0.574749532989323\\
-0.90909090909091	-0.608001644384565\\
-0.555555555555557	-0.627815497206804\\
-0.202020202020201	-0.650194256530863\\
0.151515151515152	-0.676525136047648\\
0.505050505050505	-0.709313323101579\\
0.858585858585858	-0.756961951313706\\
1.21212121212121	-0.820310359450414\\
1.56565656565656	-0.853013530624708\\
1.91919191919192	-0.888371246857336\\
2.27272727272727	-0.92691614621663\\
2.62626262626263	-0.968046484847754\\
2.97979797979798	-1.01321931130067\\
3.33333333333333	-1.0617255102895\\
3.68686868686869	-1.11512537180182\\
4.04040404040404	-1.17204301767367\\
4.39393939393939	-1.23418248469008\\
4.74747474747475	-1.30248296704874\\
5.1010101010101	-1.37557350720414\\
5.45454545454545	-1.4551261567652\\
5.80808080808081	-1.54228458950412\\
6.16161616161616	-1.63764949240515\\
6.51515151515152	-1.74381161741023\\
6.86868686868687	-1.86015238535369\\
7.22222222222222	-1.98275791545235\\
7.57575757575757	-2.11815903167507\\
7.92929292929293	-2.26640153903866\\
8.28282828282828	-2.42805836492554\\
8.63636363636364	-2.60380065290427\\
8.98989898989899	-2.78489141894691\\
9.34343434343434	-3.00086945871263\\
9.6969696969697	-3.21395878975745\\
10.0505050505051	-3.47108329972234\\
10.7575757575758	-4.1249387366083\\
11.4646464646465	-4.92081875395237\\
11.8181818181818	-5.22184874961636\\
12.1717171717172	-5.69897000433602\\
};
\addlegendentry{IC detector (uncoded system)}

\addplot [color=mycolor4, line width=1.5pt]
  table[row sep=crcr]{%
-20	-0.302374914519742\\
-15.0505050505051	-0.305808235571988\\
-11.5151515151515	-0.313413775783541\\
-9.39393939393939	-0.322933514390289\\
-8.33333333333334	-0.330444183713492\\
-6.91919191919192	-0.346455619051564\\
-6.21212121212121	-0.357344590428639\\
-5.50505050505051	-0.371811402543941\\
-4.7979797979798	-0.390975776342827\\
-4.09090909090909	-0.415707230492174\\
-3.73737373737374	-0.430750192252322\\
-3.03030303030303	-0.470515747180308\\
-2.67676767676768	-0.494017517571404\\
-2.32323232323232	-0.523288623097713\\
-1.96969696969697	-0.556350623261917\\
-1.61616161616162	-0.594875711072852\\
-1.26262626262626	-0.640073496041385\\
-0.90909090909091	-0.694317660905917\\
-0.555555555555557	-0.758460990129667\\
-0.202020202020201	-0.830934239623165\\
0.151515151515152	-0.913494242420136\\
0.505050505050505	-1.01342978930403\\
0.858585858585858	-1.12699345953268\\
1.21212121212121	-1.25238624164172\\
1.56565656565656	-1.4038209077571\\
1.91919191919192	-1.57418872801881\\
2.27272727272727	-1.76260708473827\\
2.62626262626263	-1.97077860574607\\
2.97979797979798	-2.20155653964981\\
3.33333333333333	-2.44769689066165\\
3.68686868686869	-2.70114692359029\\
4.04040404040404	-3.00392634551472\\
4.39393939393939	-3.33441900898205\\
4.74747474747475	-3.62342304294349\\
5.1010101010101	-4.03621217265444\\
5.45454545454545	-4.53760200210104\\
5.80808080808081	-5.15490195998574\\
};
\addlegendentry{M0}

\addplot [color=mycolor5, line width=1.5pt]
  table[row sep=crcr]{%
-20	-0.302195467305992\\
-14.6969696969697	-0.306648605603115\\
-11.8686868686869	-0.313428075427343\\
-10.1010101010101	-0.320863851001743\\
-8.68686868686869	-0.329899864176095\\
-7.62626262626263	-0.340862807106266\\
-6.56565656565657	-0.356040284695453\\
-5.85858585858586	-0.369603655695517\\
-5.15151515151515	-0.387787457252507\\
-4.44444444444445	-0.410872875552716\\
-3.73737373737374	-0.442301558149293\\
-3.38383838383838	-0.461862539026566\\
-3.03030303030303	-0.484134100772238\\
-2.67676767676768	-0.509686078676605\\
-2.32323232323232	-0.539857193225664\\
-1.96969696969697	-0.574756058238464\\
-1.61616161616162	-0.615806137283716\\
-1.26262626262626	-0.663078198566321\\
-0.90909090909091	-0.719612870560585\\
-0.555555555555557	-0.785235603331962\\
-0.202020202020201	-0.862934036661617\\
0.151515151515152	-0.950018755387024\\
0.505050505050505	-1.05132790425203\\
0.858585858585858	-1.16489945521226\\
1.21212121212121	-1.29364072253314\\
1.56565656565656	-1.4466996583477\\
1.91919191919192	-1.61439372640169\\
2.27272727272727	-1.81109962682409\\
2.62626262626263	-2.01304921214149\\
2.97979797979798	-2.23515164280659\\
3.33333333333333	-2.47521455067878\\
3.68686868686869	-2.72955409198204\\
4.04040404040404	-3.03621217265444\\
4.39393939393939	-3.33068311943389\\
4.74747474747475	-3.66554624884907\\
5.1010101010101	-4.13076828026902\\
5.45454545454545	-4.61978875828839\\
5.80808080808081	-5.39794000867204\\
};
\addlegendentry{M1}

\end{axis}
\end{tikzpicture}%
}
  \hfil
  \subfloat[\label{fig: I = 5dB}]{
%
%
\definecolor{mycolor1}{rgb}{0.00000,0.44700,0.74100}%
\definecolor{mycolor2}{rgb}{0.85000,0.32500,0.09800}%
\definecolor{mycolor3}{rgb}{0.92900,0.69400,0.12500}%
\definecolor{mycolor4}{rgb}{0.49400,0.18400,0.55600}%
\definecolor{mycolor5}{rgb}{0.46600,0.67400,0.18800}%
\begin{tikzpicture}[trim axis left, scale = 0.45]

\begin{axis}[%
width=4.521in,
height=3.566in,
at={(0.758in,0.481in)},
scale only axis,
xmin=-5,
xmax=3,
xlabel style={font=\color{white!15!black}},
xlabel={$\mathrm{SNR}(\si{\decibel})/\mathrm{INR}(\si{\decibel})$},
ymin=-6,
ymax=0,
ylabel style={font=\color{white!15!black}},
ylabel={$\log_{10}(\mathrm{BER})$},
axis background/.style={fill=white},
xmajorgrids,
ymajorgrids,
legend style={at={(0.03,0.03)}, anchor=south west, legend cell align=left, align=left, draw=white!15!black}
]
\addplot [color=mycolor1, line width=1.5pt]
  table[row sep=crcr]{%
-5	-0.310974546086553\\
-3.86868686868687	-0.320353194195837\\
-3.14141414141414	-0.33055945146266\\
-2.49494949494949	-0.344985529755566\\
-2.01010101010101	-0.360248632045859\\
-1.60606060606061	-0.377082761577872\\
-1.2020202020202	-0.400315570322148\\
-0.959595959595959	-0.418050341626683\\
-0.717171717171717	-0.439493195295798\\
-0.474747474747475	-0.466197448046976\\
-0.313131313131313	-0.487599802170658\\
-0.151515151515151	-0.512631099394428\\
0.0101010101010104	-0.542364828389881\\
0.171717171717172	-0.577387182212378\\
0.333333333333333	-0.619181171225503\\
0.414141414141414	-0.643358531647443\\
0.494949494949495	-0.670364188383141\\
0.575757575757576	-0.700037508186884\\
0.656565656565657	-0.732483320579169\\
0.737373737373737	-0.768726702016121\\
0.818181818181818	-0.808870493607267\\
0.898989898989899	-0.853512269318403\\
0.97979797979798	-0.902773935637858\\
1.06060606060606	-0.958666540854809\\
1.14141414141414	-1.02075218457689\\
1.22222222222222	-1.08985603893549\\
1.3030303030303	-1.16692304411026\\
1.38383838383838	-1.25494083080259\\
1.46464646464646	-1.35221617723802\\
1.54545454545455	-1.46186631274933\\
1.62626262626263	-1.58490974232889\\
1.70707070707071	-1.72544961509047\\
1.78787878787879	-1.88362473882029\\
1.86868686868687	-2.05814087347463\\
1.94949494949495	-2.25602013475816\\
2.03030303030303	-2.47860037188462\\
2.11111111111111	-2.72262002533275\\
2.19191919191919	-2.99956592252068\\
2.27272727272727	-3.32330639037513\\
2.35353535353535	-3.66154350639539\\
2.43434343434343	-4.0268721464003\\
2.51515151515152	-4.52287874528034\\
2.5959595959596	-5.22184874961636\\
2.67676767676768	-5.69897000433602\\
};
\addlegendentry{TIN detector (uncoded system)}

\addplot [color=mycolor2, line width=1.5pt]
  table[row sep=crcr]{%
-5	-0.317582333653705\\
-4.27272727272727	-0.326563438738869\\
-3.62626262626263	-0.338663501624731\\
-3.06060606060606	-0.353669806155452\\
-2.65656565656566	-0.368172841117353\\
-2.25252525252525	-0.386886839905464\\
-1.92929292929293	-0.405908134095017\\
-1.60606060606061	-0.428595164266092\\
-1.36363636363636	-0.448866078453179\\
-1.12121212121212	-0.472080236658253\\
-0.878787878787879	-0.499149029872828\\
-0.555555555555555	-0.540188499627639\\
-0.232323232323233	-0.587138104688066\\
0.0909090909090908	-0.638693445068928\\
0.414141414141414	-0.694936822491844\\
0.494949494949495	-0.710727743065982\\
0.575757575757576	-0.72907672370143\\
0.656565656565657	-0.750469014813191\\
0.737373737373737	-0.775687118607117\\
0.818181818181818	-0.808378543030345\\
0.898989898989899	-0.853512269318403\\
0.97979797979798	-0.902773935637858\\
1.06060606060606	-0.958666540854809\\
1.14141414141414	-1.02075218457689\\
1.22222222222222	-1.08985603893549\\
1.3030303030303	-1.16692304411026\\
1.38383838383838	-1.25494083080259\\
1.46464646464646	-1.35221617723802\\
1.54545454545455	-1.46186631274933\\
1.62626262626263	-1.58490974232889\\
1.70707070707071	-1.72544961509047\\
1.78787878787879	-1.88362473882029\\
1.86868686868687	-2.05814087347463\\
1.94949494949495	-2.25602013475816\\
2.03030303030303	-2.47860037188462\\
2.11111111111111	-2.72262002533275\\
2.19191919191919	-2.99956592252068\\
2.27272727272727	-3.32330639037513\\
2.35353535353535	-3.66154350639539\\
2.43434343434343	-4.0268721464003\\
2.51515151515152	-4.52287874528034\\
2.5959595959596	-5.22184874961636\\
2.67676767676768	-5.69897000433602\\
};
\addlegendentry{ML detector (uncoded system)}

\addplot [color=mycolor3, line width=1.5pt]
  table[row sep=crcr]{%
-5	-0.31677729724282\\
-4.27272727272727	-0.325300487281138\\
-3.62626262626263	-0.337393253545868\\
-3.06060606060606	-0.351928964164479\\
-2.57575757575758	-0.36952432470012\\
-2.17171717171717	-0.388515672790295\\
-1.76767676767677	-0.413206848031228\\
-1.44444444444444	-0.43753345248506\\
-1.12121212121212	-0.467064808750462\\
-0.797979797979798	-0.501809410777462\\
-0.555555555555555	-0.532489566167966\\
-0.232323232323233	-0.577272310764119\\
0.0909090909090908	-0.626896655177893\\
0.494949494949495	-0.695724949522872\\
0.656565656565657	-0.730858466287279\\
0.737373737373737	-0.752456333245515\\
0.818181818181818	-0.778979805530279\\
0.898989898989899	-0.81508438349941\\
0.97979797979798	-0.875609906579035\\
1.06060606060606	-0.958666540854809\\
1.14141414141414	-1.02075218457689\\
1.22222222222222	-1.08985603893549\\
1.3030303030303	-1.16692304411026\\
1.38383838383838	-1.25494083080259\\
1.46464646464646	-1.35221617723802\\
1.54545454545455	-1.46186631274933\\
1.62626262626263	-1.58490974232889\\
1.70707070707071	-1.72544961509047\\
1.78787878787879	-1.88362473882029\\
1.86868686868687	-2.05814087347463\\
1.94949494949495	-2.25602013475816\\
2.03030303030303	-2.47860037188462\\
2.11111111111111	-2.72262002533275\\
2.19191919191919	-2.99956592252068\\
2.27272727272727	-3.32330639037513\\
2.35353535353535	-3.66154350639539\\
2.43434343434343	-4.0268721464003\\
2.51515151515152	-4.52287874528034\\
2.5959595959596	-5.22184874961636\\
2.67676767676768	-5.69897000433602\\
};
\addlegendentry{IC detector (uncoded system)}

\addplot [color=mycolor4, line width=1.5pt]
  table[row sep=crcr]{%
-5	-0.3009813574057\\
-2.01010101010101	-0.308685059952679\\
-1.60606060606061	-0.314356772348904\\
-1.2020202020202	-0.325052559842821\\
-0.878787878787879	-0.340085700566615\\
-0.717171717171717	-0.351073336171584\\
-0.555555555555555	-0.366011712262035\\
-0.393939393939394	-0.386399167460182\\
-0.313131313131313	-0.398802939961548\\
-0.151515151515151	-0.432027221198129\\
-0.0707070707070709	-0.453879498209809\\
0.0101010101010104	-0.479356665812747\\
0.0909090909090908	-0.50970995300968\\
0.171717171717172	-0.549084560875907\\
0.252525252525253	-0.596049448287286\\
0.333333333333333	-0.653146506030556\\
0.414141414141414	-0.724260322136635\\
0.494949494949495	-0.811676295154562\\
0.575757575757576	-0.914613081174525\\
0.656565656565657	-1.03971477671937\\
0.737373737373737	-1.19485270219959\\
0.818181818181818	-1.37655428747623\\
0.898989898989899	-1.58479286409157\\
0.97979797979798	-1.81488299885741\\
1.06060606060606	-2.0945279380753\\
1.14141414141414	-2.42864060724616\\
1.22222222222222	-2.77391588402418\\
1.3030303030303	-3.14874165128092\\
1.38383838383838	-3.57511836336893\\
1.54545454545455	-4.72124639904717\\
1.62626262626263	-5.52287874528034\\
};
\addlegendentry{M0}

\addplot [color=mycolor5, line width=1.5pt]
  table[row sep=crcr]{%
-5	-0.301573203341934\\
-3.70707070707071	-0.303129228984651\\
-2.81818181818182	-0.307699596004819\\
-2.09090909090909	-0.317470458822903\\
-1.76767676767677	-0.326055241028401\\
-1.44444444444444	-0.339837737766535\\
-1.2020202020202	-0.355978178619222\\
-1.04040404040404	-0.369692157366315\\
-0.878787878787879	-0.388427500120842\\
-0.717171717171717	-0.412061286491747\\
-0.636363636363637	-0.425495319878007\\
-0.555555555555555	-0.441427017692217\\
-0.474747474747475	-0.459743122284989\\
-0.393939393939394	-0.480898186566618\\
-0.232323232323233	-0.532267616675093\\
-0.151515151515151	-0.563961454826614\\
-0.0707070707070709	-0.598461609834056\\
0.0101010101010104	-0.640885779585094\\
0.0909090909090908	-0.687000125898072\\
0.171717171717172	-0.739678129753688\\
0.252525252525253	-0.797108635496844\\
0.333333333333333	-0.865889859290268\\
0.414141414141414	-0.945712596879422\\
0.494949494949495	-1.04065804629684\\
0.575757575757576	-1.14971531688602\\
0.656565656565657	-1.27331353345499\\
0.737373737373737	-1.4191206219392\\
0.818181818181818	-1.58693505316259\\
0.898989898989899	-1.7782989356154\\
0.97979797979798	-1.99439055463972\\
1.06060606060606	-2.24680008580058\\
1.14141414141414	-2.55861911508349\\
1.22222222222222	-2.88505558428742\\
1.3030303030303	-3.24336389175415\\
1.38383838383838	-3.65364702554936\\
1.46464646464646	-4.2518119729938\\
1.62626262626263	-5.52287874528034\\
};
\addlegendentry{M1}

\end{axis}
\end{tikzpicture}%
}
  \hfil
  \subfloat[\label{fig: I = 10dB}]{
%
%
\definecolor{mycolor1}{rgb}{0.00000,0.44700,0.74100}%
\definecolor{mycolor2}{rgb}{0.85000,0.32500,0.09800}%
\definecolor{mycolor3}{rgb}{0.92900,0.69400,0.12500}%
\definecolor{mycolor4}{rgb}{0.49400,0.18400,0.55600}%
\definecolor{mycolor5}{rgb}{0.46600,0.67400,0.18800}%
\begin{tikzpicture}[trim axis left, scale = 0.45]

\begin{axis}[%
width=4.521in,
height=3.566in,
at={(0.758in,0.481in)},
scale only axis,
xmin=-3,
xmax=2,
xlabel style={font=\color{white!15!black}},
xlabel={$\mathrm{SNR}(\si{\decibel})/\mathrm{INR}(\si{\decibel})$},
ymin=-6,
ymax=0,
ylabel style={font=\color{white!15!black}},
ylabel={$\log_{10}(\mathrm{BER})$},
axis background/.style={fill=white},
xmajorgrids,
ymajorgrids,
legend style={at={(0.03,0.03)}, anchor=south west, legend cell align=left, align=left, draw=white!15!black}
]
\addplot [color=mycolor1, line width=1.5pt]
  table[row sep=crcr]{%
-3	-0.303631377791592\\
-2.24242424242424	-0.307594646983472\\
-1.78787878787879	-0.312395244514732\\
-1.38383838383838	-0.319524881989956\\
-1.08080808080808	-0.327706291699208\\
-0.828282828282829	-0.337160981721624\\
-0.626262626262626	-0.347238421412738\\
-0.474747474747475	-0.356865264463612\\
-0.323232323232324	-0.368573481343887\\
-0.171717171717172	-0.38323050472768\\
-0.0202020202020208	-0.401327654912112\\
0.0808080808080813	-0.416400000859906\\
0.181818181818182	-0.434303266553925\\
0.282828282828283	-0.456109432225015\\
0.383838383838384	-0.483228043029626\\
0.434343434343434	-0.499449310689103\\
0.484848484848485	-0.517699663261396\\
0.535353535353535	-0.53854049244223\\
0.585858585858587	-0.562947404939008\\
0.636363636363637	-0.591019312474106\\
0.686868686868687	-0.624477211711485\\
0.737373737373738	-0.663234166419431\\
0.787878787878788	-0.710937513747458\\
0.838383838383838	-0.768311278679602\\
0.888888888888889	-0.837731396761941\\
0.939393939393939	-0.922159897265006\\
0.98989898989899	-1.02553897023142\\
1.04040404040404	-1.15308792124956\\
1.09090909090909	-1.3079995901734\\
1.14141414141414	-1.49694510923198\\
1.19191919191919	-1.7307206102281\\
1.24242424242424	-2.01095084356178\\
1.29292929292929	-2.35261702988538\\
1.34343434343434	-2.7594507517174\\
1.39393939393939	-3.24336389175415\\
1.44444444444444	-3.83863199776503\\
1.49494949494949	-4.61978875828839\\
1.54545454545455	-5.69897000433602\\
};
\addlegendentry{TIN detector (uncoded system)}

\addplot [color=mycolor2, line width=1.5pt]
  table[row sep=crcr]{%
-3	-0.31198885622179\\
-2.5959595959596	-0.317943416756183\\
-2.29292929292929	-0.324543375269887\\
-1.98989898989899	-0.334038347434213\\
-1.73737373737374	-0.345428824308513\\
-1.53535353535354	-0.357439530073695\\
-1.38383838383838	-0.368594791554524\\
-1.23232323232323	-0.382050284746708\\
-1.08080808080808	-0.398823610471328\\
-0.929292929292929	-0.418740028682252\\
-0.777777777777778	-0.443355024211049\\
-0.676767676767677	-0.462314360835732\\
-0.575757575757576	-0.484126156288321\\
-0.474747474747475	-0.508739186277087\\
-0.373737373737374	-0.537013856629017\\
-0.272727272727273	-0.568604067070812\\
-0.171717171717172	-0.603798908753221\\
-0.0707070707070709	-0.642392902301619\\
0.0808080808080813	-0.705540387510317\\
0.333333333333334	-0.815552751053303\\
0.434343434343434	-0.854160555806099\\
0.484848484848485	-0.871371365460429\\
0.535353535353535	-0.884842322406985\\
0.585858585858587	-0.895775866681167\\
0.636363636363637	-0.904621400439936\\
0.686868686868687	-0.909093283198255\\
0.838383838383838	-0.916536798124378\\
0.888888888888889	-0.926340292530367\\
0.939393939393939	-0.952273545419565\\
0.98989898989899	-1.02553897023142\\
1.04040404040404	-1.15308792124956\\
1.09090909090909	-1.3079995901734\\
1.14141414141414	-1.49694510923198\\
1.19191919191919	-1.7307206102281\\
1.24242424242424	-2.01095084356178\\
1.29292929292929	-2.35261702988538\\
1.34343434343434	-2.7594507517174\\
1.39393939393939	-3.24336389175415\\
1.44444444444444	-3.83863199776503\\
1.49494949494949	-4.61978875828839\\
1.54545454545455	-5.69897000433602\\
};
\addlegendentry{ML detector (uncoded system)}

\addplot [color=mycolor3, line width=1.5pt]
  table[row sep=crcr]{%
-3	-0.311500979872945\\
-2.54545454545455	-0.318250573612815\\
-2.24242424242424	-0.325479634050299\\
-1.98989898989899	-0.33384064762037\\
-1.73737373737374	-0.345197023859543\\
-1.53535353535354	-0.357112273637483\\
-1.38383838383838	-0.36809579828816\\
-1.23232323232323	-0.381579511578735\\
-1.08080808080808	-0.398182195361513\\
-0.929292929292929	-0.417979841306108\\
-0.828282828282829	-0.433898523587198\\
-0.727272727272728	-0.451606896215265\\
-0.626262626262626	-0.471490802533173\\
-0.525252525252525	-0.49441458838845\\
-0.424242424242425	-0.520325376291485\\
-0.323232323232324	-0.550230111951308\\
-0.222222222222222	-0.583391109119636\\
-0.121212121212121	-0.620203345430629\\
-0.0202020202020208	-0.659861386879532\\
0.0808080808080813	-0.702677285794698\\
0.232323232323233	-0.768930736236749\\
0.333333333333334	-0.811904360685863\\
0.434343434343434	-0.850999629714983\\
0.484848484848485	-0.867833200490552\\
0.535353535353535	-0.881960252392609\\
0.585858585858587	-0.893265299902154\\
0.636363636363637	-0.901397802333984\\
0.686868686868687	-0.906021793961457\\
0.737373737373738	-0.908955919012036\\
0.787878787878788	-0.91032445443761\\
0.838383838383838	-0.913230986201613\\
0.888888888888889	-0.921992934912089\\
0.939393939393939	-0.946825484293668\\
0.98989898989899	-1.00985213467351\\
1.04040404040404	-1.15308792124956\\
1.09090909090909	-1.3079995901734\\
1.14141414141414	-1.49694510923198\\
1.19191919191919	-1.7307206102281\\
1.24242424242424	-2.01095084356178\\
1.29292929292929	-2.35261702988538\\
1.34343434343434	-2.7594507517174\\
1.39393939393939	-3.24336389175415\\
1.44444444444444	-3.83863199776503\\
1.49494949494949	-4.61978875828839\\
1.54545454545455	-5.69897000433602\\
};
\addlegendentry{IC detector (uncoded system)}

\addplot [color=mycolor4, line width=1.5pt]
  table[row sep=crcr]{%
-3	-0.300873677786027\\
-1.03030303030303	-0.303368441393185\\
-0.626262626262626	-0.307684601734659\\
-0.373737373737374	-0.313555005402354\\
-0.222222222222222	-0.319670833315263\\
-0.121212121212121	-0.325623004691335\\
-0.0202020202020208	-0.334676097346242\\
0.0808080808080813	-0.347153412805692\\
0.131313131313131	-0.355124381630074\\
0.181818181818182	-0.365458245423047\\
0.232323232323233	-0.378201813491362\\
0.282828282828283	-0.394169376954995\\
0.333333333333334	-0.41268423193934\\
0.383838383838384	-0.438160704358606\\
0.434343434343434	-0.469366224317599\\
0.484848484848485	-0.508082484867915\\
0.535353535353535	-0.55983268400947\\
0.585858585858587	-0.625634808387533\\
0.636363636363637	-0.713195605309314\\
0.686868686868687	-0.831384677788983\\
0.737373737373738	-0.985786179750028\\
0.787878787878788	-1.18802305566305\\
0.838383838383838	-1.44523230247192\\
0.888888888888889	-1.78264768011864\\
0.939393939393939	-2.19579239491796\\
0.98989898989899	-2.68655462957359\\
1.04040404040404	-3.2839966563652\\
1.09090909090909	-3.97469413473523\\
1.14141414141414	-4.72124639904717\\
1.19191919191919	-5.52287874528034\\
};
\addlegendentry{M0}

\addplot [color=mycolor5, line width=1.5pt]
  table[row sep=crcr]{%
-3	-0.30194471387526\\
-1.63636363636364	-0.305422000158409\\
-1.43434343434343	-0.307941339800291\\
-1.23232323232323	-0.312350665827958\\
-1.03030303030303	-0.320566652026924\\
-0.878787878787879	-0.330222100695976\\
-0.777777777777778	-0.339588017993415\\
-0.727272727272728	-0.345096070643982\\
-0.626262626262626	-0.360627079641155\\
-0.575757575757576	-0.370059581238317\\
-0.474747474747475	-0.396280919602689\\
-0.424242424242425	-0.413368818050521\\
-0.373737373737374	-0.432325618188869\\
-0.323232323232324	-0.455447057471236\\
-0.272727272727273	-0.481726962954947\\
-0.222222222222222	-0.514802597589356\\
-0.171717171717172	-0.553101071571075\\
-0.121212121212121	-0.600272643867093\\
-0.0707070707070709	-0.655403400513942\\
-0.0202020202020208	-0.719421629631924\\
0.0303030303030303	-0.7949770812459\\
0.0808080808080813	-0.878689708832511\\
0.131313131313131	-0.975616965996668\\
0.181818181818182	-1.08047737579763\\
0.232323232323233	-1.19633807676378\\
0.383838383838384	-1.56624616248324\\
0.484848484848485	-1.7966406984777\\
0.535353535353535	-1.8980468225228\\
0.585858585858587	-1.97187628487116\\
0.636363636363637	-2.03782014709123\\
0.686868686868687	-2.08618614761628\\
0.737373737373738	-2.13739163603506\\
0.787878787878788	-2.20162562331844\\
0.838383838383838	-2.2954205503037\\
0.888888888888889	-2.42562143558692\\
0.939393939393939	-2.6848696828164\\
0.98989898989899	-3.01099538430146\\
1.04040404040404	-3.52287874528034\\
1.09090909090909	-4.17392519729917\\
1.19191919191919	-5.52287874528034\\
};
\addlegendentry{M1}

\end{axis}
\end{tikzpicture}%
}
  \caption{\small BER vs SNR for BPSK modulation in AWGN+ARI channel with \protect\subref{fig: I = 1dB} $I = \SI{1}{\decibel}$, \protect\subref{fig: I = 5dB} $I = \SI{5}{\decibel}$ and \protect\subref{fig: I = 10dB} $I = \SI{10}{\decibel}$}
  \label{fig: Bit Error Rate (BER) vs S for BPSK modulation in AWGN+ARI channel with I = 1dB, I = 5dB and I = 10dB}
\end{figure*}

\begin{figure}
  \centering
  \input{"Plots/LDPC codes/BER_simulation_AWGN_channel_ARI_fixedSNR_1dB_LDPC"}
  \caption{\small BER vs INR for BPSK modulation in AWGN+ARI channel with $S = \SI{1}{\decibel}$}
  \label{fig: Bit Error Rate (BER) vs I for BPSK modulation in AWGN+ARI channel with S = 1dB (LDPC code)}
\end{figure}

\begin{figure*}
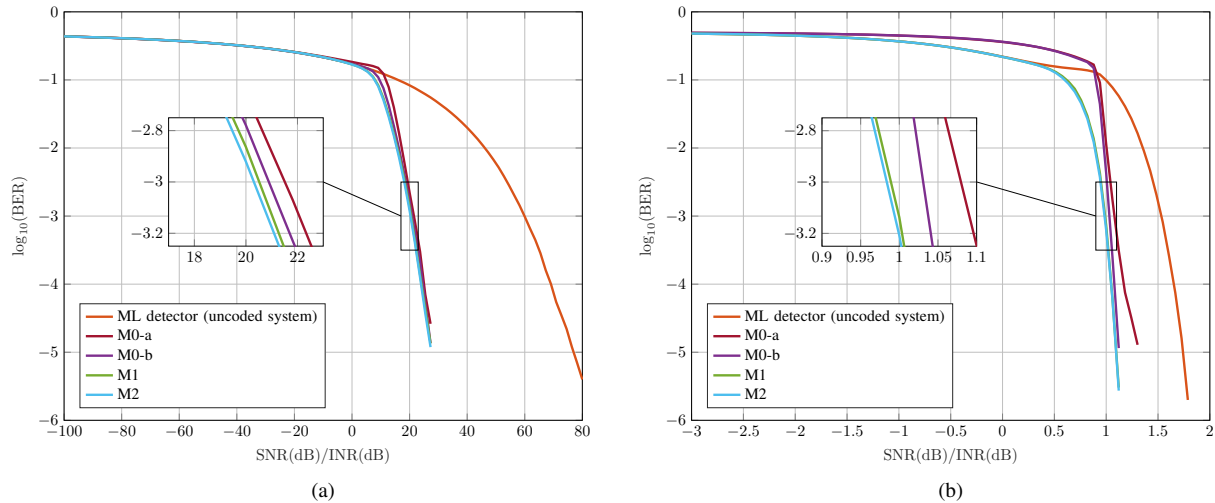

  \centering
  \subfloat[\label{fig: I = 0.15dB}]{\input{"Plots/LDPC codes/BER_simulation_AWGN_channel_ARI_fixedINR_0.15dB_LDPC"}}
  \hfil
  \subfloat[\label{fig: I = 8.25dB}]{\input{"Plots/LDPC codes/BER_simulation_AWGN_channel_ARI_fixedINR_8.25dB_LDPC"}}
  \caption{\small BER vs SNR for BPSK modulation in AWGN+ARI channel with \protect\subref{fig: I = 0.15dB} $I = \SI{0.15}{\decibel}$ and \protect\subref{fig: I = 8.25dB} $I = \SI{8.25}{\decibel}$}
  \label{fig: Bit Error Rate (BER) vs S for BPSK modulation in AWGN+ARI channel with I = 0.15dB and I = 8.25dB}
\end{figure*}

\subsection{M1: Convolutional Codes}
\label{sec: Numerical Results: M1conv}
In Fig.~\ref{fig: Bit Error Rate (BER) vs I for BPSK modulation in AWGN+ARI channel with S = 1dB (convolutional code)} 
we compare the BER vs INR for the AWGN+ARI channel with fixed $S = \SI{1}{\decibel}$ and BPSK modulation, for the following receivers:
baseline M0) the Viterbi algorithm with the standard AWGN-only channel metric, and
methodology M1) the Viterbi algorithm with the AWGN+ARI channel metric.
Note that for baseline M0 we do not make any distinction between M0-a or M0-b, since the AWGN-only metric depends in this case only on the inner product between vectors $\bm{x}$ and $\bm{y}$.
For reference, we also plot the BER of an uncoded system with the three decoders derived in~\cite{8332962}, namely
ML = Maximum Likelihood (optimal symbol-by-symbol maximum a posteriori detector), 
TIN = Treat Interference as Noise (approximately optimal in the low INR regime),  and
IC = Interference Cancellation (approximately optimal in the high INR regime).
The tested one-half rate convolutional code is defined by the octal digits $\bm{g}^{(0)} = 5$ and $\bm{g}^{(1)} = 7$ \cite[p. 540]{lin2004error}.
As already pointed out in~\cite{8332962} for uncoded systems, for coded systems we can also clearly identify two distinct regimes depending on whether we are in the low INR regime or in the high INR regime. 
When $I \ll S$, the performance of the Viterbi algorithm does not depend on the metric used; this is expected as in this regime the radar interference has negligible power.
When $I \gg S$, M1 outperforms both M0 and the uncoded schemes; this is expected as in this regime the radar interference is very powerful and thus must be appropriately dealt with. In this last case the baseline scheme gives $\mathrm{BER} \approx 1/2$, so we would not even need to use a code.

We notice that, when using in practice the new decoding metric, we need not distinguish between AWGN-only channel and AWGN+ARI channel, as the decoding metric for the AWGN+ARI channel with $I = 0$ gives the decoding metric for the AWGN-only channel. Therefore, in AWGN+ARI channels, one would need to measure both the SNR and the INR in order to appropriately tune the decoding metric, and not trivially ``collapse'' them into a single number $\mathrm{SINR} = \frac{S}{1+I}$.

Similar conclusions can be drawn if we keep the INR fixed and let the SNR vary, as one can see from Fig.~\ref{fig: Bit Error Rate (BER) vs S for BPSK modulation in AWGN+ARI channel with I = 1dB, I = 5dB and I = 10dB} where we plot the BER vs SNR for the AWGN+ARI channel with fixed values of $I$.
On the one hand, it is evident from Fig.~\ref{fig: I = 10dB} that BER curves coincide for both metrics when $I \ll S$.
On the other hand, the modified metric leads to a better performance when $I \gg S$, i.e. when we are in the high INR regime.

\subsection{M1: LDPC Codes}
\label{sec: Numerical Results: M1ldpc}

In Fig.~\ref{fig: Bit Error Rate (BER) vs I for BPSK modulation in AWGN+ARI channel with S = 1dB (LDPC code)} we again fixed $S = \SI{1}{\decibel}$ and plot the BER curves vs INR for:
baseline M0-a) the SPA for the AWGN-only channel as if INR is zero,
baseline M0-b) the SPA for the AWGN-only channel with SNR substituted with SINR,
methodology M1) the SPA for the AWGN+ARI channel, and ML detector
for an uncoded system.
We use the $(63, 37)$ EG-LDPC code, generated from the 2-D Euclidean geometry $\mathrm{EG}(2, 2^{3})$, and set the maximum number of decoding iterations to $5$, a good trade-off between the decoding speed and the performance \cite[p.~884]{lin2004error}. 
We see that for LDPC codes we can draw the same conclusions as before for the convolutional codes. In particular, we observe a slight improvement in the low INR region for baseline M0-b when the SNR parameter in baseline M0-a is replaced with the SINR term; this is reasonable, since this methodology does not neglect the radar interference.

\subsection{M2: LDPC Code Design for the AWGN+ARI Channel}
\label{sec: Numerical Results: M2ldpc}

We run the optimization algorithm for methodology M2 as described in Section~\ref{sec: Channel Model and Codes Used}. The optimized degree distributions are reported in \tablename~\ref{tab: Coefficients of lambda and rho optimized degree distributions}, for which we set $d_{\text{v}} = 30$ and $d_{\text{c}} = 11$.
For the optimization in the AWGN-only channel we choose $S = \SI{-2.53}{\decibel}$, while for the AWGN+ARI channel we consider two cases:
$I \ll S$ with $S = \SI{0.45}{\decibel}, I = \SI{0.15}{\decibel}$, and 
$I \gg S$ with $S = \SI{2.75}{\decibel}, I = \SI{8.25}{\decibel}$.

\begin{table}
  \centering\scriptsize
  \caption{Optimized degree distributions.}
  \label{tab: Coefficients of lambda and rho optimized degree distributions}
  \begin{tabular}{cccc}
    \toprule
    $\lambda_{i}$ or $\rho_{i}$ & AWGN-only & AWGN+ARI $S \gg I$ & AWGN+ARI $S \ll I$\\
    \midrule
    $\lambda_{2}$ & $0.1907$ & $0.1937$ & $0.1962$\\
    $\lambda_{3}$ & $0.0963$ & $0.0596$ & $0.0774$\\
    $\lambda_{4}$ & $0.1126$ & $0.2325$ & $0.1944$\\
    $\lambda_{5}$ & $0.1095$ & &\\
    $\lambda_{30}$ & $0.4909$ & $0.5142$ & $0.5320$\\
    \midrule
    $\rho_{10}$ & $0.5193$ & $0.5266$ & $0.4991$\\
    $\rho_{11}$ & $0.4807$ & $0.4734$ & $0.5009$\\
    \bottomrule
  \end{tabular}
\end{table}

We converted our degree distributions into parity-check matrices by means of the geometrical systematic approach presented in~\cite[Chapter 17, pp. 922-929]{lin2004error} that avoids, by construction, the presence of cycles of length $4$.
We generated a $(4032, 1984)$ LDPC code with rate $R = 0.49$, so as to have a fair comparison among AWGN-only and AWGN+ARI codes.

In Fig.~\ref{fig: Bit Error Rate (BER) vs S for BPSK modulation in AWGN+ARI channel with I = 0.15dB and I = 8.25dB} we plot the BER vs SNR curves for the following codes:
M0-a) LDPC with AWGN-only initialization as if INR is zero and code optimized for the AWGN-only channel,
M0-b) LDPC with AWGN-only initialization using SINR instead of SNR and code optimized for the AWGN-only channel
M1) LDPC with AWGN+ARI initialization and code optimized for the AWGN-only channel, and
M2) LDPC with AWGN+ARI initialization and code optimized for the AWGN+ARI channel.
For comparison we also plot the BER of the ML detector for the uncoded system.
Baseline M0-b outperforms again baseline M0-a, which shows also an error floor (Fig.~\ref{fig: I = 8.25dB}) due to numerical problems related to the implementation of the SPA; the best BER is obtained, unsurprisingly, with M2. However somewhat surprisingly, the performance gain of M2 over M1 does not appear so significant.

This last observation can be explained by keeping in mind that codes designed for AWGN-only have been previously observed to behave well for a large class of channels~\cite{910578}. We speculate that an even more accurate optimization (e.g. density evolution rather than EXIT charts) would not yield significantly better codes.
This may also be intuitively understood as follows: AWGN is known to be the ``worst noise'' in terms of capacity among all noises with the same second-moment; thus, a code designed for AWGN-only is intrinsically robust and performs quite well in non-AWGN scenarios. 

From Fig.~\ref{fig: I = 8.25dB} we observe that, for good BER performance, it is more important to use the correct decoding metric than to re-design the code. In practice, this is a good news: current wireless system specifications in terms of channel codes need not be changed when the band is shared with radar systems, and in order to effectively cope with the extra interference it suffices to use the appropriate INR-based decoding metric.

\section{Conclusion}\label{sec: Conclusion}
In this paper we studied channel codes for wireless channels with radar interference. We derived a new decoding metric for this new channel model and used it with known classes of codes. We compared two design methodologies: using codes designed for AWGN-only with the new decoding metric, and designing codes directly optimized for the radar interfered channel. We found that the BER improvement of the latter over the former is not so significant, meaning that in future wireless systems it is much more critical to use the correct decoder than to re-design codes. This work focused on binary codes mapped onto the BPSK modulation; extensions to other modulation schemes, and possibly other classes of codes (such as Polar codes) is an interesting and unexplored research direction.

The work of the Authors was supported in part by NSF award number 1443967.
The Authors would like to thank Prof. Roberto Garello for initial discussion on LDPC codes.

\bibliographystyle{IEEEtran}
\bibliography{references}

\begin{thebibliography}{1}
\providecommand{\url}[1]{#1}
\csname url@samestyle\endcsname
\providecommand{\newblock}{\relax}
\providecommand{\bibinfo}[2]{#2}
\providecommand{\BIBentrySTDinterwordspacing}{\spaceskip=0pt\relax}
\providecommand{\BIBentryALTinterwordstretchfactor}{4}
\providecommand{\BIBentryALTinterwordspacing}{\spaceskip=\fontdimen2\font plus
\BIBentryALTinterwordstretchfactor\fontdimen3\font minus
  \fontdimen4\font\relax}
\providecommand{\BIBforeignlanguage}[2]{{%
\expandafter\ifx\csname l@#1\endcsname\relax
\typeout{** WARNING: IEEEtran.bst: No hyphenation pattern has been}%
\typeout{** loaded for the language `#1'. Using the pattern for}%
\typeout{** the default language instead.}%
\else
\language=\csname l@#1\endcsname
\fi
#2}}
\providecommand{\BIBdecl}{\relax}
\BIBdecl

\bibitem{7852330}
S.~{Shahi}, D.~{Tuninetti}, and N.~{Devroye}, ``On the capacity of the awgn
  channel with additive radar interference,'' in \emph{2016 54th Annual
  Allerton Conference on Communication, Control, and Computing (Allerton)},
  Sep. 2016, pp. 902--907.

\bibitem{8332962}
N.~{Nartasilpa}, A.~{Salim}, D.~{Tuninetti}, and N.~{Devroye}, ``Communications
  system performance and design in the presence of radar interference,''
  \emph{IEEE Trans. on Comm.}, vol.~66, no.~9, pp. 4170--4185, Sep. 2018.

\bibitem{910578}
T.~J. {Richardson}, M.~A. {Shokrollahi}, and R.~L. {Urbanke}, ``Design of
  capacity-approaching irregular low-density parity-check codes,'' \emph{IEEE
  Trans. on Info. Theory}, vol.~47, no.~2, pp. 619--637, Feb 2001.

\bibitem{910580}
S.-Y. {Chung}, T.~J. {Richardson}, and R.~L. {Urbanke}, ``Analysis of
  sum-product decoding of low-density parity-check codes using a gaussian
  approximation,'' \emph{IEEE Trans. on Info. Theory}, vol.~47, no.~2, pp.
  657--670, Feb 2001.

\bibitem{richardson_urbanke_2008}
T.~Richardson and R.~Urbanke, \emph{Modern Coding Theory}.\hskip 1em plus 0.5em
  minus 0.4em\relax Cambridge University Press, 2008.

\bibitem{lin2004error}
S.~Lin and D.~J. Costello, \emph{Error control coding: fundamentals and
  applications}.\hskip 1em plus 0.5em minus 0.4em\relax Upper Saddle River, NJ:
  Pearson/Prentice Hall, 2004.

\bibitem{Amraoui:85786}
\BIBentryALTinterwordspacing
A.~Amraoui, ``Asymptotic and finite-length optimization of ldpc codes,'' Ph.D.
  dissertation, EPFL, Lausanne, 2006. [Online]. Available:
  \url{http://infoscience.epfl.ch/record/85786}
\BIBentrySTDinterwordspacing

\end{thebibliography}

\end{document}